\documentstyle[amssymb,prl,aps,epsfig]{revtex}

\begin{document}

\title{Finite size effect in Bi$_{2}$Sr$_{2}$CaCu$_{2}$O$_{8+\delta }$ and  YBa$_{2}
$Cu$_{3}$O$_{6.7}$ probed by the in-plane and out of plane
penetration depths}
\author{T. Schneider and D. Di Castro\\
Physik-Institut der Universit\"{a}t Z\"{u}rich, Winterthurerstrasse 190,\\
CH-8057 Z\"{u}rich, Switzerland}
\maketitle
\begin{abstract}
We report on a systematic finite size scaling analysis of in-plane
penetration depth data taken on
Bi$_{2}$Sr$_{2}$CaCu$_{2}$O$_{8+\delta }$ epitaxially-grown films
and single crystals, and of in-plane and out of plane data taken
on YBa$_{2}$Cu$_{3}$O$_{6.7}$ aligned powder. It is shown that the
tails in temperature dependence of the penetration depths,
appearing around the transition temperature, are fully consistent
with a finite size effect. This uncovers the granular nature of
these cuprates, consisting of superconducting homogeneous domains
of nanoscale extent, embedded in a non-superconducting matrix.
\end{abstract}

\bigskip

Since the discovery of superconductivity in cuprates by Bednorz
and M\"{u}ller\cite{bed} a tremendous amount of work has been
devoted to their characterization. The issue of inhomogeneity and
its characterization is essential for several reasons, including:
First, if inhomogeneity is an intrinsic property, a
re-interpretation of experiments, measuring an average of the
electronic properties, is unavoidable. Second, inhomogeneity may
point to a microscopic phase separation, i.e. superconducting
grains, embedded in a non superconducting matrix. Third, there is
neutron spectroscopic evidence for nanoscale cluster formation and
percolative superconductivity in various
cuprates\cite{mesot,furrer}. Fourth, nanoscale spatial variations
in the electronic characteristics have been observed in underdoped
Bi$_{2}$Sr$_{2}$CaCu$_{2}$O$_{8+\delta }$ with scanning tunnelling
microscopy (STM)\cite{liu,chang,cren,lang}. They reveal a spatial
segregation of the electronic structure into 3nm diameter
superconducting domains in an electronically distinct background.
On the contrary, a large degree of homogeneity has been observed
by Renner and Fischer\cite{renner}. As STM is a surface probe, the
relevance of these observations for bulk and thermodynamic
properties has been clarified in terms of a finite size scaling of
specific heat and London penetration depth data, revealing $\
L_{ab}\approx 50\AA$ and $L_{c}\approx 68\AA$ for the length scale
of the superconducting domains along the c-axis and in the
ab-plane, respectively \cite{tsfsbi}. Fifth, superconducting domains 
with $\
L_{ab}\approx 300\AA$ were revealed by x-ray diffraction 
 in oxygen doped 
La$_{2}$CuO$_{4}$ single crystal \cite{daniele2}. Sixth, 
in
YBa$_{2}$Cu$_{3}$O$_{7-\delta }$, MgB$_{2}$, 2H-NbSe$_{2}$ and
Nb$_{77}$Zr$_{23}$ considerably larger domains have been
established. The magnetic field induced finite size effect
revealed lower bounds ranging from $L=182$ to $814$\AA
\cite{tsfs}.

This paper addresses these issues by performing a detailed finite
size scaling analysis of the tails in the measured temperature
dependence of the penetration depths, appearing around the
transition temperature. Specifically we analyze the data taken on
Bi$_{2}$Sr$_{2}$CaCu$_{2}$O$_{8+\delta }$ in the form of\
epitaxially-grown films\cite{osborn} and high quality single
crystals\cite{jacobs,daniele}, and on magnetically aligned
YBa$_{2}$Cu$_{3}$O$_{6.7}$ powder\cite{panagop}. The paper is
organized as follows. Next we sketch the finite size scaling
theory adapted for the analysis of penetration depth data. Then we
analyze the experimental data and establish the consistency with a
finite size effect, uncovering the granular nature of these
cuprates, consisting of homogeneous superconducting domains with
nanoscale extent, embedded in a non-superconducting matrix.

Supposing that cuprate superconductors are granular, consisting of
spatial superconducting domains, embedded in a non-superconducting
matrix and with spatial extent $L_{a}$, $L_{b}$ and $L_{c}$ along
the crystallographic a, b and c-axis. in this case the correlation
length $\xi _{i}$ in direction $i$, increasing strongly when
$T_{c}$ is approached cannot grow beyond $L_{i}$. Consequently,
for finite superconducting domains, the thermodynamic quantities,
like the specific heat and penetration depth, are smooth functions
of temperature. As a remnant of the singularity at $T_{c}$ these
quantities exhibit a so called finite size effect\cite{fisher},
namely a maximum or an inflection point at $T_{p_{i}}$, where $\xi
_{i}\left( T_{p_{i}}\right) =L_{i}$. There is mounting
experimental evidence that, for the accessible temperature ranges,
the effective finite temperature critical behavior of the cuprates
is controlled by the critical point of uncharged superfluids
(3D-XY)\cite{osborn,book}. In this case there is the universal
relationship
\begin{equation}
\frac{1}{\lambda _{i}^{2}\left( T\right) }=\frac{16\pi
^{3}k_{B}T}{\Phi _{0}^{2}\xi _{i}^{t}\left( T\right) },
\label{eq1}
\end{equation}
between the London penetration depth $\lambda _{i}$ and the
transverse correlation length $\xi _{i}^{t}$ in direction
$i$\cite{book}. As aforementioned, when the superconductor is
inhomogeneous, consisting of superconducting grains with length
scales $L_{i}$, embedded in a non-superconducting matrix,\ the
$\xi _{i}^{t}$'s do not diverge but are bounded by
\begin{equation}
\xi _{i}^{t}\xi _{j}^{t}\leq L_{k}^{2},\ i\neq j\neq k.
\label{eq2}
\end{equation}
A characteristic feature of the resulting finite size effect is
the occurrence of an inflection point at $T_{p_{k}}$ below
$T_{c}$, the transition temperature of the homogeneous system.
Here
\begin{equation}
\xi _{i}^{t}\left( T_{p_{k}}\right) \xi _{j}^{t}\left(
T_{p_{k}}\right) =L_{k}^{2},\ i\neq j\neq k,  \label{eq3}
\end{equation}
and Eq.(\ref{eq1}) reduces to
\begin{equation}
\left. \frac{1}{\lambda _{i}\left( T\right) \lambda _{j}\left(
T\right) }\right| _{T=T_{p_{k}}}=\frac{16\pi
^{3}k_{B}T_{p_{k}}}{\Phi _{0}^{2}}\frac{1}{L_{k}}.  \label{eq4}
\end{equation}
In the homogeneous case $1/\left( \lambda _{i}\left( T\right)
\lambda _{j}\left( T\right) \right) $ decreases continuously with
increasing temperature and vanishes at $T_{c}$, while for
superconducting domains, embedded in a non-superconducting matrix,
it does not vanish and exhibits a turning point at
$T_{p_{k}}<T_{c}$, so that
\begin{equation}
\left. d\left( \frac{1}{\lambda _{i}\left( T\right) \lambda
_{j}\left( T\right) }\right) /dT\right| _{T=T_{p_{k}}}=0
\label{eq5}
\end{equation}
Since the experimental data for the temperature dependence of the
penetration depths is available in the form $\lambda _{ab}$ and
$\lambda _{c} $ only, we rewrite Eq.(\ref{eq4}) as
\begin{equation}
L_{c}=\frac{16\pi ^{3}k_{B}T_{p_{c}}\left( \lambda _{a}\left(
T\right) \lambda _{b}\left( T\right) \right) _{T=T_{p_{c}}}}{\Phi
_{0}^{2}}\approx \frac{16\pi ^{3}k_{B}T_{p_{c}}\lambda
_{ab}^{2}\left( T_{p_{c}}\right) }{\Phi _{0}^{2}},  \label{eq6}
\end{equation}
and
\begin{equation}
L_{b}=\frac{16\pi ^{3}k_{B}T_{p_{b}}\left( \lambda _{a}\left(
T\right) \lambda _{c}\left( T\right) \right) _{T=T_{p_{b}}}}{\Phi
_{0}^{2}}\approx L_{ab}=\frac{16\pi ^{3}k_{B}T_{p_{b}}\left(
\lambda _{ab}\left( T\right) \lambda _{c}\left( T\right) \right)
_{T=T_{p_{ab}}}}{\Phi _{0}^{2}} \label{eq7}
\end{equation}
to derive estimates for the diameter of the superconducting grains
along the c-axis and parallel to the ab-plane. Noting that in the
homogeneous system the transverse correlation lengths diverge as

\begin{equation}
\xi _{i}^{t}\left( T\right) =\xi _{0i}^{t}\left| t\right| ^{-\nu
},\ t=\frac{T}{T_{c}}-1,\ \nu \approx 2/3,  \label{eq8}
\end{equation}
the critical amplitudes and associated critical properties of the
homogeneous system can also be derived from a finite size
analysis. From Eqs.(\ref{eq6}) and (\ref{eq7}) we obtain the
relations
\begin{equation}
\xi _{ab}^{t}\left( T_{p_{c}}\right) =L_{c},\ \sqrt{\xi
_{ab}^{t}\left( T_{p_{ab}}\right) \xi _{c}^{t}\left(
T_{p_{ab}}\right) }=L_{ab}.  \label{eq9}
\end{equation}
Indeed, the transverse correlation lengths cannot grow beyond the
corresponding limiting length scales. Thus given $T_{p_{c}}$and
$L_{c}$, $T_{p_{ab}}$ and $L_{ab}$, the critical amplitudes $\xi
_{0ab}^{t}$ and $\xi _{0c}^{t}$ are readily derived. The
anisotropy follows then from
\begin{equation}
\gamma =\left( \frac{\xi _{0c}^{t}}{\xi _{0ab}^{t}}\right)
^{1/2}=\frac{\lambda _{0c}}{\lambda _{0ab}}.  \label{eq10}
\end{equation}

As the system feels its finite size when the correlation length
becomes of the order of the confining length, a physical quantity
$O$ adopts the scaling form\cite{schultka}
\begin{equation}
\frac{O\left( t,L\right) }{O\left( t,L=\infty \right) }=f\left(
x\right) ,\ x=\frac{L}{\xi \left( t,L=\infty \right) }.
\label{eq10a}
\end{equation}
The point is that the scaling function $f$ depends only on the
dimensionless ratio $L/\xi \left( t,L=\infty \right) $ and it does
not depend on microscopic details of the system. It does, however,
depend on the observable $O$, the type of confining geometry and on
the conditions imposed (or not, in the case of free boundaries) at
the boundaries of the domains. Considering the London penetration
depth, it reduces to
\begin{equation}
\frac{\lambda _{0i}\lambda _{0j}}{\lambda _{i}\lambda _{j}}\left|
t\right| ^{-\nu }=f\left( \frac{sign\left( t\right) L_{k}\left|
t\right| ^{\nu }}{\sqrt{\xi _{0i}^{t}\xi _{0j}^{t}}}\right)
=g\left( y\right) ,\ \ y=sign\left( t\right) \left| t\right|
\left( \frac{L_{k}}{\sqrt{\xi _{0i}^{t}\xi _{0j}^{t}}}\right)
^{1/\nu }=sign\left( t\right) \left| \frac{t}{t_{p_{k}}}\right|
\label{eq10b}
\end{equation}
For $t$ small and $L_{k}\rightarrow \infty $, $\pm y\rightarrow
\infty $ and
\begin{equation}
g\left( y\rightarrow -\infty \right) =1,\ g\left( y\rightarrow
\infty \right) =0  \label{eq10c}
\end{equation}
while for $t=0$ and $L_{k}\neq 0$
\begin{equation}
g\left( y\rightarrow 0\right) =g_{0k}y^{-\nu }=g_{0k}\left( \left|
\frac{t}{t_{p_{k}}}\right| \right) ^{-\nu },  \label{eq10d}
\end{equation}
so that in this limit
\begin{equation}
\frac{\lambda _{0i}\lambda _{0j}}{\lambda _{i}\left(
T_{c},L_{k}\right) \lambda _{j}\left( T_{c},L_{k}\right)
}=g_{0k}\frac{\sqrt{\xi _{0i}^{t}\xi _{0j}^{t}}}{L_{k}}.
\label{eq10e}
\end{equation}
As expected, a sharp superconductor to normal state transition
requires domains of infinite extent. Moreover at $t_{p}$, $y=1$
and $d\left( \lambda _{0i}\lambda _{0j}/\left( \lambda _{i}\left(
T\right) \lambda _{j}\left( T\right) \right) \right) /dt=0$.
Accordingly, there is an inflection point at $T_{p_{k}}$. Since
the scaling function $g\left( y\right) $ depends on the type of
confining geometry and on the conditions imposed (or not, in the
case of free boundaries) at the boundaries of the domains, this
applies to the amplitude $g_{0k}$ as well.

\begin{figure}[tbp]
\centering
\includegraphics[totalheight=6cm]{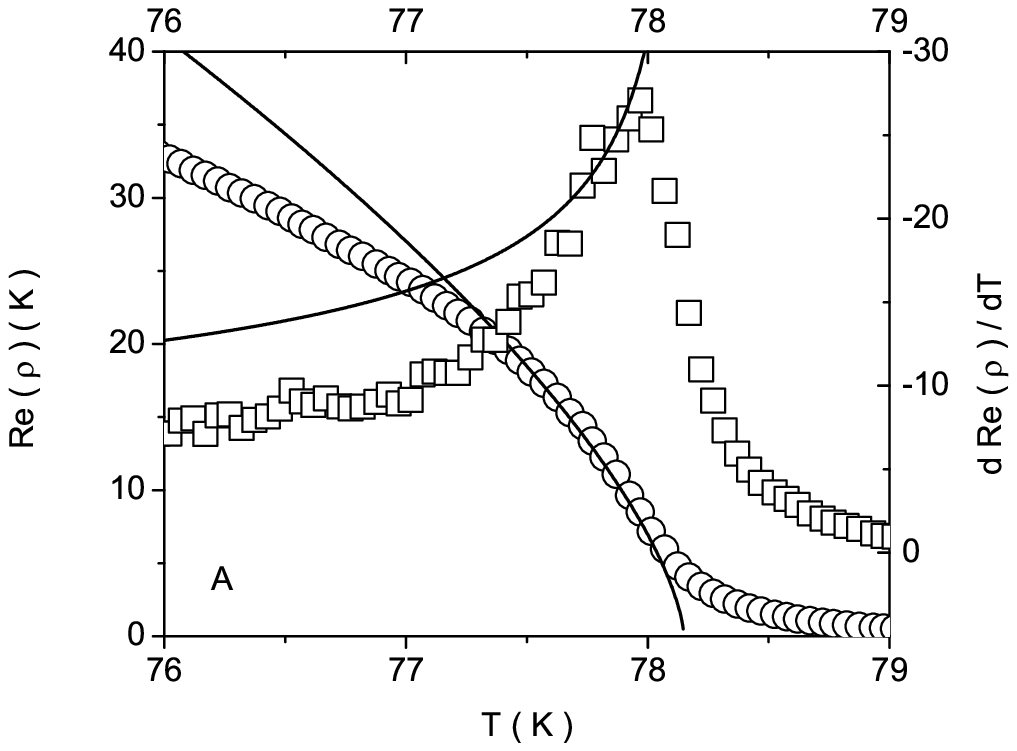}
\includegraphics[totalheight=6cm]{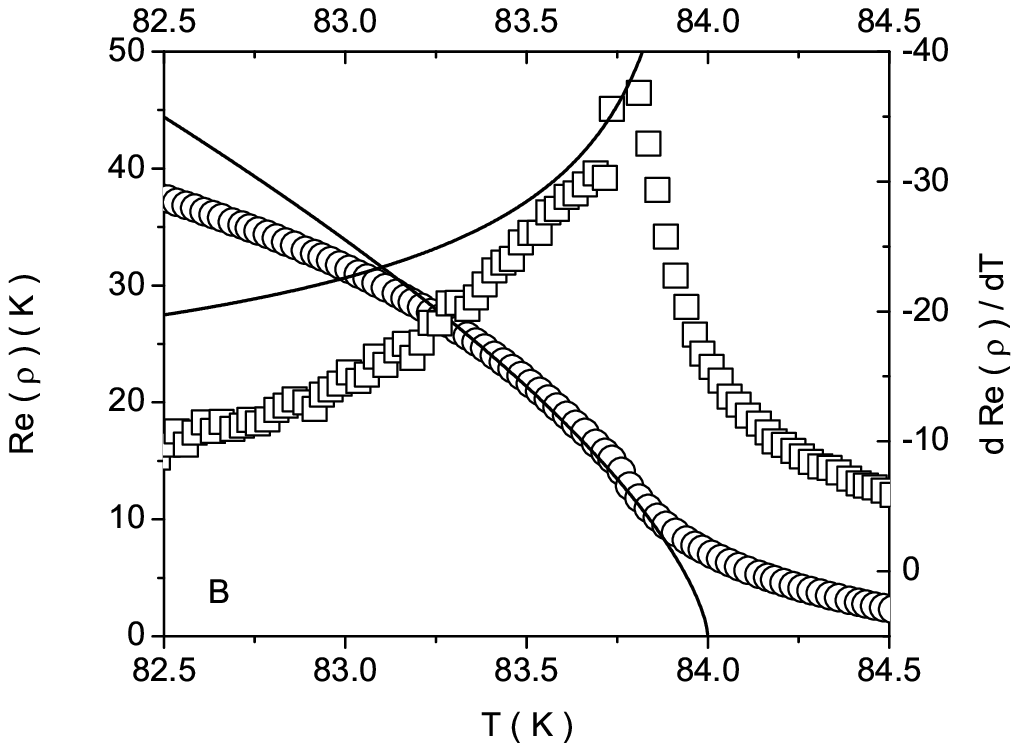}
\includegraphics[totalheight=6cm]{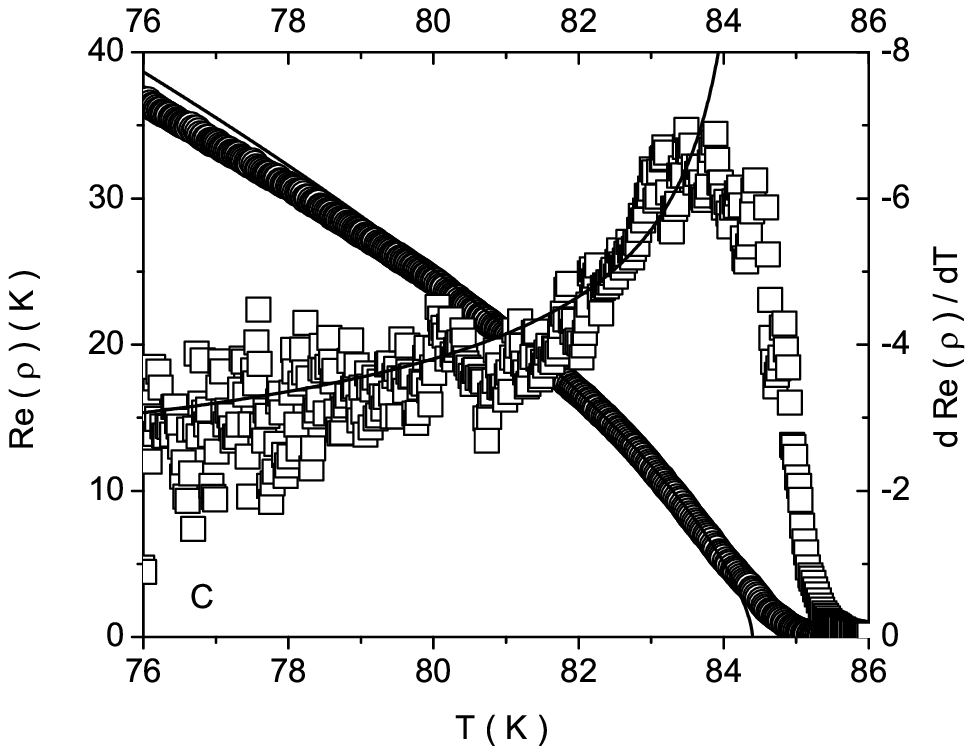}
\caption{$Re\left( \rho \right) \propto 1/\lambda _{ab}^{2}$ and
$dRe\left( \rho \right) /dT$ versus $T$ for films A, B and C taken
from Osborn \emph{\ et al.}\protect\cite{osborn}. The solid lines
indicate the leading critical behavior of a homogeneous bulk
system according to Eq.(\ref{eq11}) with the parameters listed in
Table I.} \label{fig1}
\end{figure}

We are now prepared to analyze the extended and systematic
penetration depth data of Osborn \emph{\ et al.}\cite{osborn}
derived from complex conductivity measurements on
epitaxially-grown Bi$_{2}$Sr$_{2}$CaCu$_{2}$O$_{8+\delta }$ films
using a two-coil inductive technique at zero applied field. In
Fig.\ref {fig1} we displayed the temperature dependence of the
real part of the complex superfluid density which is proportional
to $1/\lambda _{ab}^{2}$ for films A, B and C with different
doping level. For comparison we included the leading critical
behavior in terms of
\begin{equation}
Re\left( \rho \right) =Re\left( \rho \right) _{0}\left| t\right|
^{\nu },\ \nu \simeq 2/3  \label{eq11}
\end{equation}
with the parameters listed in Table I. Apparently the data is
inconsistent with such a sharp transition. It clearly uncovers a
rounded transition which occurs smoothly and with that a finite
size effect at work. Indeed the extreme in $d\Re\left( \rho
\right) /dT$ clearly reveals the existence of an inflection point
at $T_{p}$ below the bulk $T_{c}$, where the transverse
correlation length $\xi _{ab}^{t}$ attains the limiting length
along the c-axis (Eq.(\ref{eq9})). Using the estimates for
$T_{p}$, derived from the location of the extremum in $d\Re\left(
\rho \right) /dT$ values for $L_{c}$ are now readily calculated
with the aid of Eq.(\ref{eq6}) and the parameters listed in Table
I. The results included in this table clearly point to the
nanoscale nature of the inhomogeneities along the c-axis.
Nevertheless, due to the small critical amplitude of the
transverse correlation length $\xi _{0ab}^{t}$, which follows from
Eq.(\ref{eq3}) and the corresponding parameters listed in Table I,
and the fact that $L_{c}$ is considerably smaller than the film
thickness $d$, the critical 3D-XY critical regime is attained.

\bigskip
\begin{center}
\begin{tabular}{|c|c|c|c|c|c|c|}
\hline & A & B & C & SC1 & SC1 & SC2 \\ \hline $T_{c}$ (K) & 78.15
& 84 & 84.4 & 87.5 & 87.5 & 91 \\ \hline $T_{p_{c}}$ (K) & 77.97 &
83.77 & 83.73 & 87 & 87 & 90.5 \\ \hline $-t_{p_{c}}$ & 0.0023 &
0.0027 & 0.0079 & 0.0057 & 0.0057 & 0.0055 \\ \hline $Re\left(
\rho \right) _{0}$ ( K ) & 450 & 650 & 180 & - & - &  \\ \hline
$\left( \lambda _{ab}\left( 0\right) /\lambda _{0ab}\right) ^{2}$
& - & - & - & 1.25 & 1.25 & 1.65 \\ \hline
$\lambda _{ab}\left( 0\right) $ (nm) & 235 & 265 & 285 & 180 & 250 & 185 \\
\hline $Re\left( \rho \left( 0\right) \right) $ ( K ) & 169.3 &
135.1 & 116.9 & - & - &  \\ \hline $Re\left( \rho \left(
T_{p_{c}}\right) \right) $ ( K ) & 8.51 & 13.66 & 7.1 & - & - &
\\ \hline $\left( \lambda _{ab}\left( 0\right) /\lambda
_{ab}\left( T_{p_{c}}\right) \right) ^{2}$ & - & - & - & 0.066 &
0.066 & 0.062 \\ \hline $L_{c}$ (\AA) & 137 & 93 & 180 & 68 & 132
& 80 \\ \hline $\xi _{0ab}^{t}$ (A) & 2.39 & 1.8 & 7.14 & 2.17 &
4.21 & 2.5 \\ \hline $d$ ( \AA ) & 323 & 616 & 924 & - & - &  \\
\hline $\left( \lambda _{0ab}\left( 0\right) /\lambda _{ab}\left(
T_{c}\right) \right) ^{2}$ & - & - & - & 0.045 & 0.045 & 0.038 \\
\hline $Re\left( \rho \left( T_{c}\right) \right) /Re\left( \rho
\right) _{0}$ & 0.097 & 0.011 & 0.018 & - & - &  \\ \hline g$_{0}$
& 0.6 & 0.6 & 0.6 & 1.6 & 1.6 & 1.2 \\ \hline $g_{0}\xi
_{0ab}^{t}/L_{c}$ & 0.010 & 0.011 & 0.024 & 0.051 & 0.051 & 0.038
\\ \hline
\end{tabular}
\end{center}

\bigskip
Table I: Estimates entering and resulting from the finite size
scaling analysis of the in-plane penetration depth data of
Bi$_{2}$Sr$_{2}$CaCu$_{2}$O$_{8+\delta }$ films (A,B,C) and single
crystals (SC1 and SC2). The films are respectively, A overdoped, B
slightly overdoped and C underdoped.

\bigskip

\begin{figure}[tbp]
\centering
\includegraphics[totalheight=6cm]{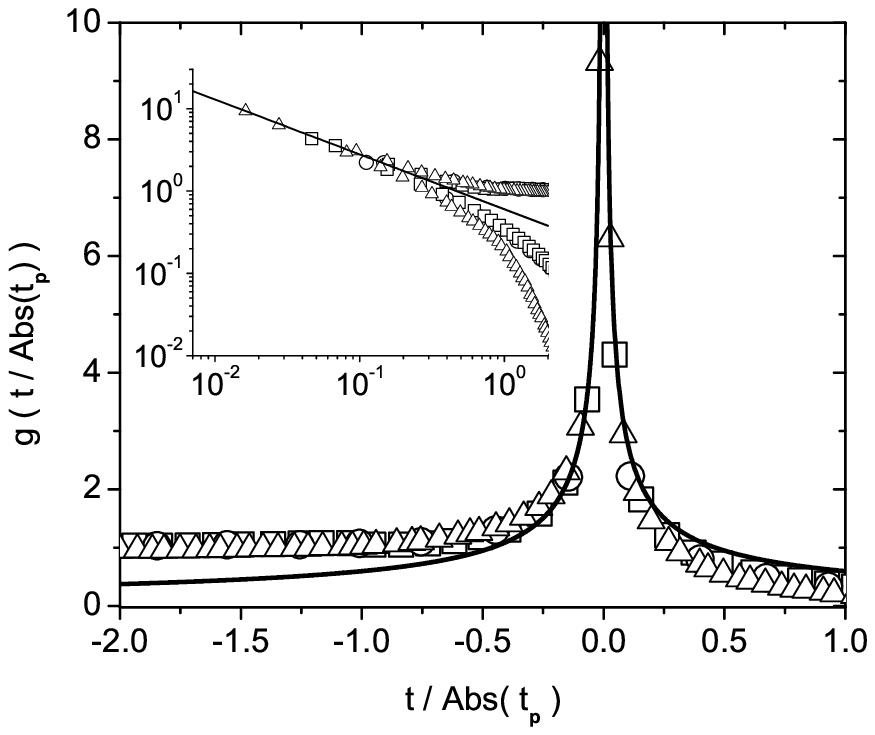}
\caption{Finite size scaling function $g$ versus $t/\left|
t_{p_{c}}\right| $ for the films A $\left( \bigcirc \right) $, B
$\left( \square \right) $ and C $\left( \Delta \right) $ derived
from the data of Osborn \emph{\ etal.} \protect\cite{osborn}. The
insert displays the log plot and the solid lines are
Eq.(\ref{eq13}).} \label{fig2}
\end{figure}

To extend the analysis further we consider the finite size scaling
function entering Eq.(\ref{eq10b}) in terms of
\begin{equation}
g\left( \frac{t}{\left| t_{p_{c}}\right| }\right) =\left| t\right|
^{-\nu }\frac{Re\left( \rho \right) }{Re\left( \rho \right) _{0}}.
\label{eq12}
\end{equation}
In Fig.\ref{fig2} we displayed this finite size scaling function
for films A, B and C. The collapse of the data on a single curve
indicates that the inhomogeneities, also differing in their extent
$L_{c}$, have nearly the same shape and are subject to the same
boundary conditions. Indeed close to $t=0$ the data approaches the
expected behavior (Eq.(\ref{eq10b}))
\begin{equation}
g\left( \frac{t}{\left| t_{p_{c}}\right| }\right) =g_{0c}\left|
\frac{t}{t_{p_{c}}}\right| ^{-2/3},  \label{eq13}
\end{equation}
and $g_{0c}\simeq 0.6$. Moreover rewriting Eq.(\ref{eq10e}) in the
form
\begin{equation}
\frac{Re\left( \rho \left( T_{c}\right) \right) }{Re\left( \rho
\right) _{0}}=g_{0c}\left| t_{p_{c}}\right| ^{\nu
}=\frac{g_{0c}\xi _{0ab}^{t}}{L_{c}}, \label{eq14}
\end{equation}
this relation provides a consistency test of the estimates for
$g_{0c}$, $\xi _{0ab}^{t}$ and $L_{c}$. From Table I it is seen
that there is satisfactory agreement.
\bigskip

\begin{figure}[tbp]
\centering
\includegraphics[totalheight=6cm]{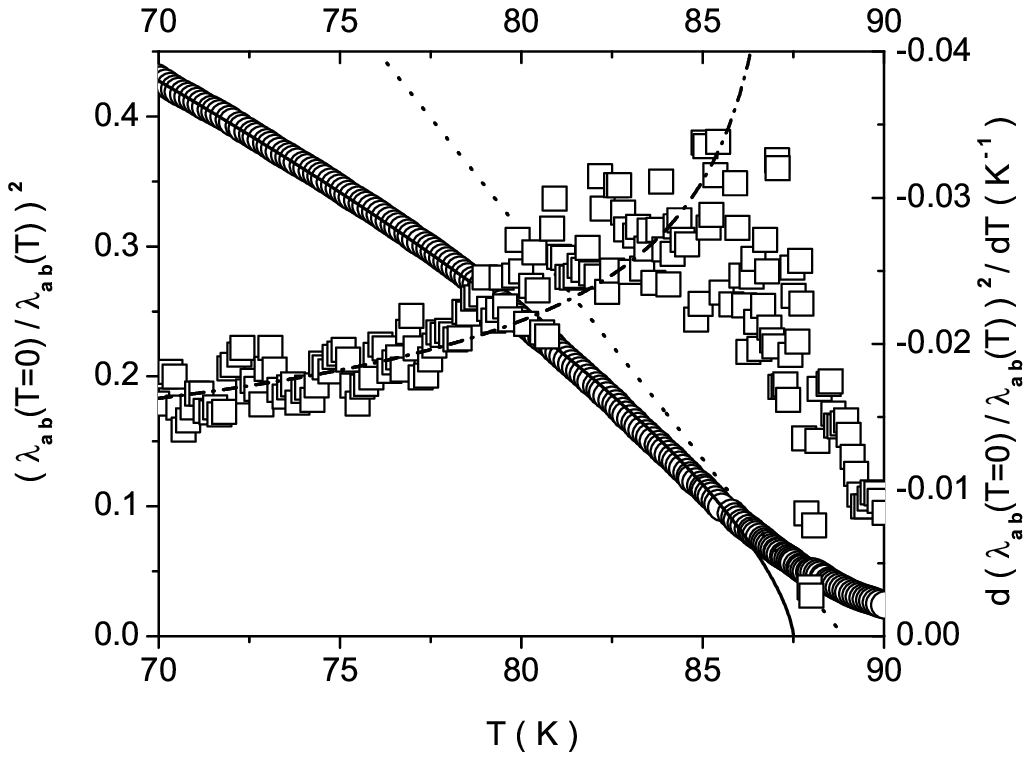}
\caption{Microwave surface impedance data for $\lambda
_{ab}^{2}\left( T=0\right) /\lambda _{ab}^{2}\left( T\right) $ and
 $d\left( \lambda _{ab}^{2}\left(
T=0\right) \right/\lambda _{ab}^{2}\left( T\right) ) /dT$ versus
$T$ of a high-quality Bi$_{2}$Sr$_{2}$CaCu$_{2}$O$_{8+\delta }$
single crystal taken from Jacobs \emph{et al.}
\protect\cite{jacobs}. The solid line is $\lambda _{ab}^{2}\left(
T=0\right) /\lambda _{ab}^{2}\left( T\right) =1.25\left(
1-T/T_{c}\right) ^{2/3}$ with $T_{c}=87.5$K, the dash-dot line its
derivative, indicating the behavior of the homogeneous bulk
system, and the dashed line is the tangent at the inflection
point, $T_{p_{c}}\approx 87$K.} \label{fig3}
\end{figure}

Next we turn to the microwave surface impedance data for the
in-plane penetration depth of a high-quality
Bi$_{2}$Sr$_{2}$CaCu$_{2}$O$_{8+\delta }$ single crystal. In
Fig.\ref{fig3} we displayed the data of Jacobs \emph{et al.}
\cite{jacobs}. In analogy to the film data shown in Fig.\ref{fig1}
there is clear evidence for a rounded transition, giving rise to
an inflection point. Using Eq.(\ref{eq6}) the finite size scaling
analysis yields the estimates listed in Table I (SC1). In this
context it should be kept in mind that there is still a
considerable uncertainty in the absolute value of the zero
temperature in-plane penetration depth $\lambda _{ab}\left(
0\right) $, the estimates ranging from $1800$ to $2690$ \AA\cite
{prozorov}. For this reason we considered in Table I $\lambda
_{ab}\left( 0\right) =1800$ and $2500$ \AA, leading to
$L_{c}\simeq 68$ and $132$ \AA, respectively. In any case, due to
the small critical amplitude of the transverse correlation length
$\xi _{0ab}^{t}$, the critical 3D-XY regime is attainable in both
cases and in agreement with Fig.\ref{fig3}.

\begin{figure}[tbp]
\centering
\includegraphics[totalheight=6cm]{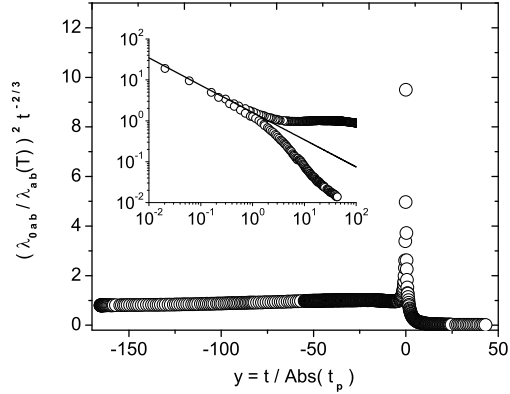}
\caption{Finite size scaling function $g$ versus $t/\left|
t_{p_{c}}\right| $ for the in-plane penetration depth data shown
in Fig.\ref{fig3}, derived from the data of Jacobs \emph{et
al.}\protect\cite{jacobs}.The insert shows the log plot and the
solid lines are Eq.(\ref{eq13}) with $g_{0c}=1.6$.} \label{fig4}
\end{figure}

To substantiate the occurrence of a finite size effect and to
clarify whether or not the rather small $L_{c}$ value is
associated with a different shape of the superconducting domains
and different boundary conditions, we derived and display in
Fig.\ref{fig4} the finite size scaling function. The parameters
entering this derivation are listed in Table I. Although the curve
adopts the generic shape, in analogy to the film data shown in
Fig.\ref {fig3}, there is an essential difference. Indeed,
$g_{0c}=1.6$, entering Eq.(\ref{eq13}), differs substantially from
the film value $g_{0c}=0.6$.

\begin{figure}[tbp]
\centering
\includegraphics[totalheight=6cm]{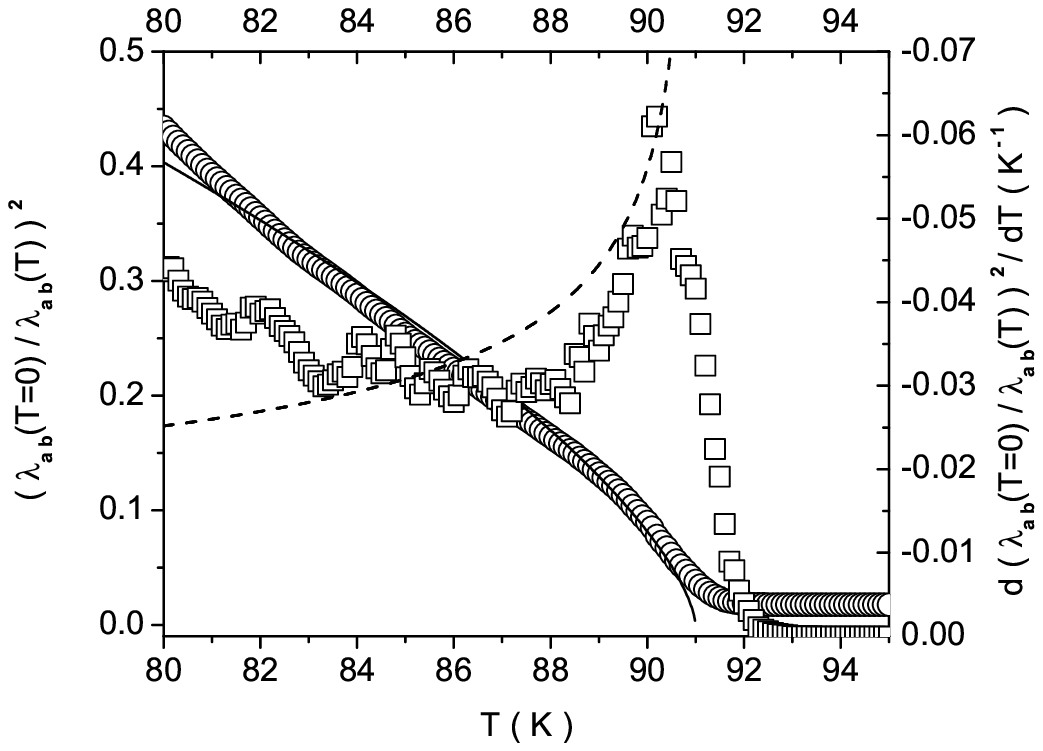}
\caption{Single coil inductance data for $\lambda _{ab}^{2}\left(
T=0\right) /\lambda _{ab}^{2}\left( T\right) $ and $d\left(
\lambda _{ab}^{2}\left( T=0\right) /\lambda _{ab}^{2}\left(
T\right) \right) /dT$ versus $T$ of a high-quality
Bi$_{2}$Sr$_{2}$CaCu$_{2}$O$_{8+\delta }$ single crystal taken
from Di Castro \emph{et al.} \protect\cite{daniele}. The solid
line is $\lambda _{ab}^{2}\left( T=0\right) /\lambda
_{ab}^{2}\left( T\right) =1.9\left( 1-T/T_{c}\right) ^{2/3}$with
$T_{c}=91$K, the dashed line its derivative, indicating the
behavior of the homogeneous bulk system.} \label{fig5}
\end{figure}

\begin{figure}[tbp]
\centering
\includegraphics[totalheight=6cm]{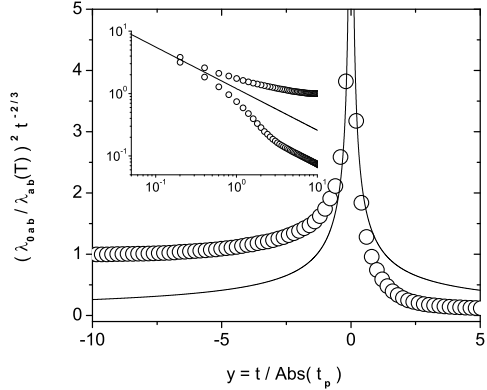}
\caption{Finite size scaling function $g$ versus $t/\left|
t_{p_{c}}\right| $ for the in-plane penetration depth data shown
in Fig.\ref{fig5}. The insert shows the log plot and the solid
lines are Eq.(\ref{eq13}) with $g_{0c}=1.2$.} \label{fig6}
\end{figure}

To provide some hints concerning the extrinsic or intrinsic nature
of the inhomogeneities, we consider next the data taken on a high
quality optimally doped Bi$_{2}$Sr$_{2}$CaCu$_{2}$O$_{8+\delta }$
single crystal of different provenance. In Fig.\ref{fig5} we
displayed the data for the sample with $T_{c}=91 $K, obtained with
the single coil inductance method\cite {daniele}. In analogy to
the data shown in Figs.\ref{fig1} and \ref{fig3} there is clear
evidence for a rounded transition, giving rise to an inflection
point. Invoking the finite size scaling analysis outlined above
and the parameters listed in Table I (SC2) we obtain with
Eq.(\ref{eq6}) for the limiting length along the c-axis the
estimate $L_{c}\simeq 80$\AA. Although the data is not very dense
around $T_{c}$, the resulting scaling function shown in
Fig.\ref{fig6} is consistent with the expected behavior of the
finite size scaling function. The solid lines are Eq.(\ref{eq13})
with $g_{0c}=1.2$.

Summarizing the in-plane penetration depth data taken on three
Bi$_{2}$Sr$_{2}$CaCu$_{2}$O$_{8+\delta }$ films and two single
crystals, we uncovered in all these samples clear evidence for a
finite size effect, stemming from the tail in the temperature
dependence of the in-plane penetration depth. We have shown that
the scaling behavior of this tail is fully consistent with a
finite size effect, arising from the finite extent of the
superconducting domains along the c-axis (see Figs.\ref{fig2},
\ref{fig4} and \ref{fig6}). Noting that in these five samples
$L_{c}$ is of nanoscale and varies from $68$ to $137$ \AA only, it
is suggestive to trace the occurrence of domains back to an
intrinsic phenomenon. Furthermore, from Table I it is seen that
the $g_{0c}$ value of the films differs substantially from that of
the single crystals. Since $g_{0c}$ depends on the shape of the
superconducting domains and the boundary conditions at their
surface, it becomes clear that the inhomogeneities in the films
differ in an essential manner from those in single crystals.
Moreover, given the estimates for $g_{0c}$, $\xi _{0ab}^{t}$and
$L_{c}$, $g_{0c}\xi _{0ab}^{t}/L_{c}$ is readily calculated. The
resulting estimates provide an independent check of the
consistency of the finite size scaling analysis. Indeed, according
to Eq.(\ref{eq10e}), this quantity should equal $\left( \lambda
_{0ab}\left( 0\right) /\lambda _{ab}\left( T_{c}\right) \right)
^{2}$. In Table I we observe satisfactory consistency.

\begin{figure}[tbp]
\centering
\includegraphics[totalheight=6cm]{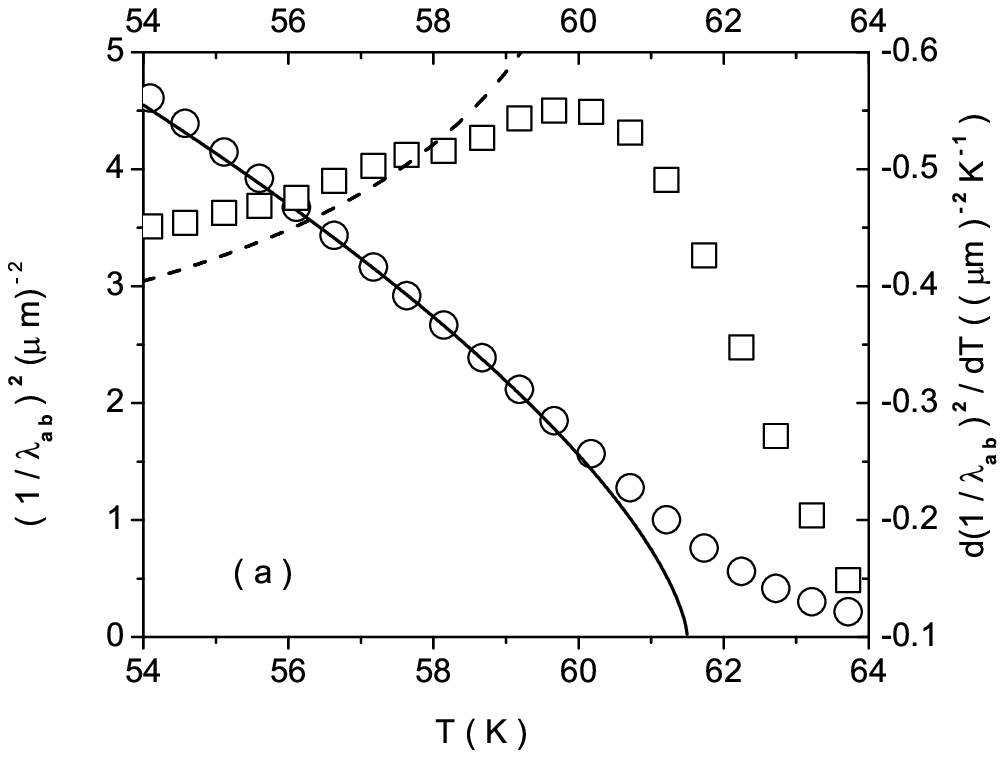}
\includegraphics[totalheight=6cm]{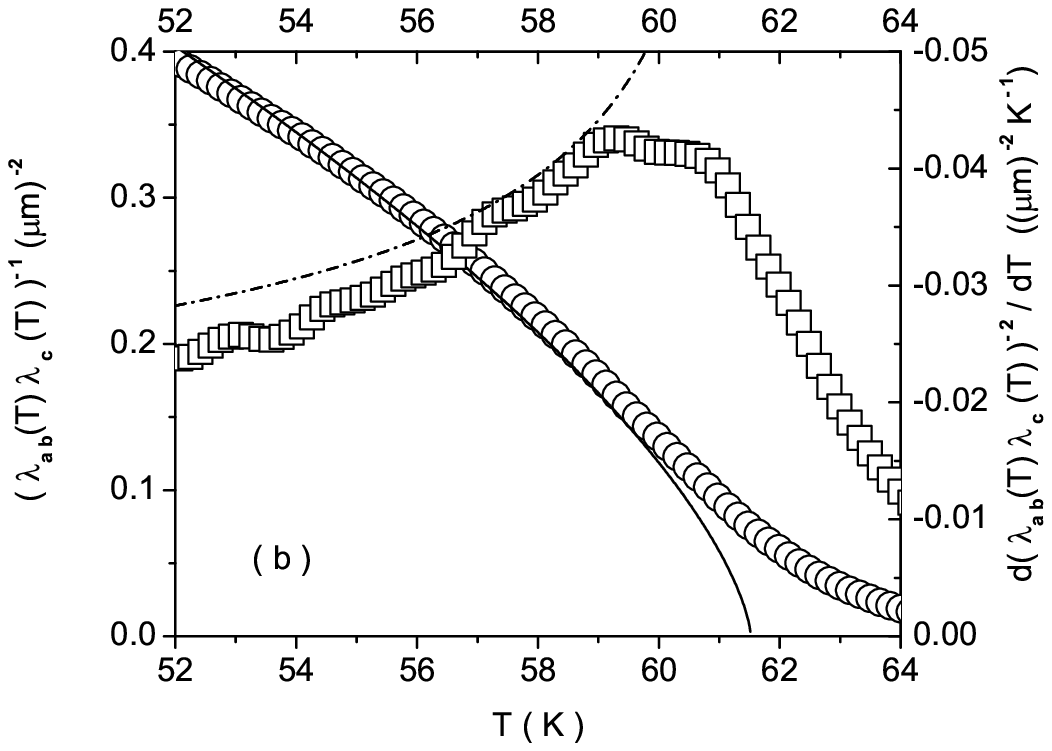}
\caption{(a) $\left( 1/\lambda _{ab}\left( T\right) \right) ^{2}$
$\left( \bigcirc \right) $ and $d\left( 1/\lambda _{ab}\left(
T\right) \right) ^{2}/dT$ $\left( \square \right) $ versus $T$ for
YBa$_{2}$Cu$_{3}$O$_{6.7}$ derived from the data of Panagopoulos
\emph{et al}.\protect\cite{panagop}. The solid lines is $\left(
1/\lambda _{ab}\left( T\right) \right) ^{2}=\left( 1/\lambda
_{0ab}\right) ^{2}\left| t\right| ^{2/3}$ and the dashed line its
derivative, indicating the leading critical behavior of the
homogeneous system. The corresponding values for $T_{c}$ and the
critical amplitude $\left( 1/\lambda _{0ab}\right) ^{2}$ are
listed in Table II; (b) $1/\lambda _{ab}\left( T\right) /\lambda
_{c}\left( T\right) $ $\left( \bigcirc \right) $ and $d\left(
1/\lambda _{ab}\left( T\right) /\lambda _{c}\left( T\right)
\right) /dT$ $\left( \square \right) $ versus $T$ for
YBa$_{2}$Cu$_{3}$O$_{6.7}$ derived from the data of Panagopoulos
\emph{et al}.\protect\cite{panagop}. The solid line is $1/\lambda
_{ab}\left( T\right) /\lambda _{c}\left( T\right) =1/\left(
\lambda _{0ab}\lambda _{0c}\right) \left| t\right| ^{2/3}$ and the
dashed one its derivative, indicating the leading critical
behavior of the homogeneous system. The corresponding values for
$T_{c}$ and the critical amplitude $1/\left( \lambda _{0ab}\lambda
_{0c}\right) $ are listed in Table II, while the parameters
resulting from the finite size scaling analysis are summarized in
Table III.} \label{fig7}
\end{figure}

To substantiate the generic occurrence of the finite size effect
we turn to the data taken on magnetic field c-axis aligned
YBa$_{2}$Cu$_{3}$O$_{6.7}$ powder, where both, the temperature
dependence of the in-plane and out of plane penetration depths,
have been measured\cite {panagop}. In Fig.\ref{fig7} we displayed
respectively, $\left( 1/\lambda _{ab}\left( T\right) \right)
^{2}$, $d\left( 1/\lambda _{ab}\left( T\right) \right) ^{2}/dT$
versus $T$ and $d\left( 1/\lambda _{ab}\left( T\right) /\lambda
_{c}\left( T\right) \right) /dT$ versus $T$. The solid lines are
respectively, $\left( 1/\lambda _{ab}\left( T\right) \right)
^{2}=\left( 1/\lambda _{0ab}\right) ^{2}\left| t\right| ^{2/3}$
and $1/\lambda _{ab}\left( T\right) /\lambda _{c}\left( T\right)
=1/\left( \lambda _{0ab}\lambda _{0c}\right) \left| t\right|
^{2/3}$, while the dashed lines are the respective derivatives.
These lines indicate the leading critical behavior of the
homogeneous system. The corresponding values for $T_{c}$ and the
critical amplitude $1/\left( \lambda _{0ab}\lambda _{0c}\right) $
are listed in Table II, while the parameters of the finite size
scaling analysis are summarized in Table III. Here we used
Eqs.(\ref{eq6}) and (\ref{eq7}) to obtain the estimates for
$L_{c}$ and $L_{ab}$, the diameters of the superconducting
domains. While $L_{c}$ is comparable to the values found in
Bi$_{2}$Sr$_{2}$CaCu$_{2}$O$_{8+\delta }$ (see Table I), $L_{ab}$
turns out to be an order of magnitude larger. Its value
$L_{ab}=592$\AA  is consistent with the lower bound
$L_{ab}>576$\AA, derived from the magnetic field induced finite
size effect on the specific heat of
YBa$_{2}$Cu$_{3}$O$_{6.6}$\cite {tsfs}.

\begin{center}
\begin{tabular}{|c|c|c|c|c|c|c|c|c|}
\hline $T_{c}$ & $\left( \lambda _{0ab}\right) ^{-2}$ & $\left(
\lambda _{0ab}\lambda _{0c}\right) ^{-1}$ & $\left( \lambda
_{0c}\right) ^{-2}$ & $\gamma $ & $\left( \lambda _{ab}\left(
T_{c}\right) \right) ^{-2}$ & $\left( \lambda _{ab}\left(
T_{c}\right) \lambda _{c}\left( T_{c}\right) \right) ^{-1}$ &
$\left( \frac{\lambda _{0ab}}{\lambda _{ab}\left( T_{c}\right) }
\right) ^{2}$ & $\frac{\lambda _{0ab}\lambda _{0c}}{\lambda
_{ab}\left( T_{c}\right) \lambda _{c}\left( T_{c}\right) }$ \\
\hline (K) & $\left( \mu \text{m}\right) ^{-2}$ & $\left( \mu
\text{m}\right) ^{-2}$ & $\left( \mu \text{m}\right) ^{-2}$ &  &
$\left( \mu \text{m}\right) ^{-2}$ & $\left( \mu \text{m}\right)
^{-2}$ &  &  \\ \hline 61.5 & 18.5 & 1.4 & 0.11 & 13.2 & 0.874 &
0.076 & 0.047 & 0.054 \\ \hline
\end{tabular}
\end{center}
\bigskip
Table II: Estimates for $T_{c}$, $\left( \lambda _{0ab}\right)
^{-2}$, $\left( \lambda _{0ab}\lambda _{0c}\right) ^{-1}$,
$\lambda _{0c}$, $\gamma =\lambda _{0c}/\lambda _{0ab}$, $\lambda
_{ab}\left( T_{c}\right) $ and $\lambda _{ab}\left( T_{c}\right)
\lambda _{c}\left( T_{c}\right) $ for YBa$_{2}$Cu$_{3}$O$_{6.7}$
derived from the data shown in Figs.\ref{fig7}.

\bigskip

\begin{center}
\begin{tabular}{|l|l|l|l|l|l|l|l|l|l|l|l|}
\hline $T_{p_{c}}$ & $T_{p_{ab}}$ & $\left( \frac{1}{\lambda
_{ab}\left( T_{p_{c}}\right) }\right) _{T_{p_{c}}}^{2}$ & $\left(
\frac{1}{\lambda _{ab}\left( T\right) \lambda _{c}\left( T\right)
}\right) _{T_{p_{ab}}}$ & $L_{c}$ & $L_{ab}$ & $\xi _{0ab}^{t}$ &
$\sqrt{\xi _{0ab}^{t}\xi _{0c}^{t}}$ & $g_{0c}$ & $\frac{g_{0c}\xi
_{0ab}^{t}}{L_{c}}$ & $g_{0ab}$ & $\frac{g_{0ab}\sqrt{\xi
_{0ab}^{t}\xi _{0c}^{t}}}{L_{ab}}$ \\ \hline (K) & (K) & $\left(
\mu \text{m}\right) ^{-2}$ & $\left( \mu
\text{m}\right) ^{-2}$ & (\AA) & (\AA) & (\AA) & (\AA) &  &  &  &  \\
\hline 59.66 & 59.3 & 1.85 & 0.16 & 52 & 592 & 5 & 64.3 & 0.5 &
0.05 & 0.55 & 0.06
\\ \hline
\end{tabular}
\end{center}

\bigskip

Table III: Finite size scaling estimates for
YBa$_{2}$Cu$_{3}$O$_{6.7}$ derived from the data shown in
Figs.\ref{fig7}.

(( (a) $\left| t\right| ^{-\nu }\left( \lambda
_{0ab}/\lambda _{ab}\left( T\right) \right) ^{2}$ versus $t/\left|
t_{p_{c}}\right| $ and (b) $\left| t\right| ^{-\nu }\left( \left(
\lambda _{0ab}\lambda _{0c}\right) /\left( \lambda _{ab}\left(
T\right) \lambda _{c}\left( T\right) \right) \right) $ versus
$t/\left| t_{p_{ab}}\right| $ derived from the data shown in
Fig.\ref{fig7} for YBa$_{2}$Cu$_{3}$O$_{6.7}.$ ))

\bigskip

\begin{figure}[tbp]
\centering
\includegraphics[totalheight=6cm]{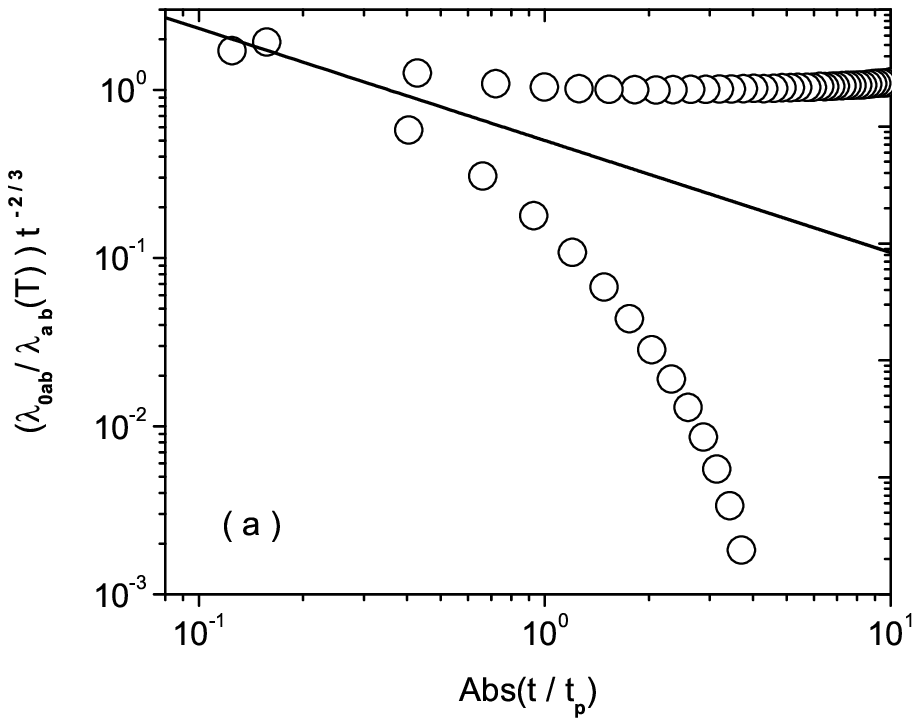}
\includegraphics[totalheight=6cm]{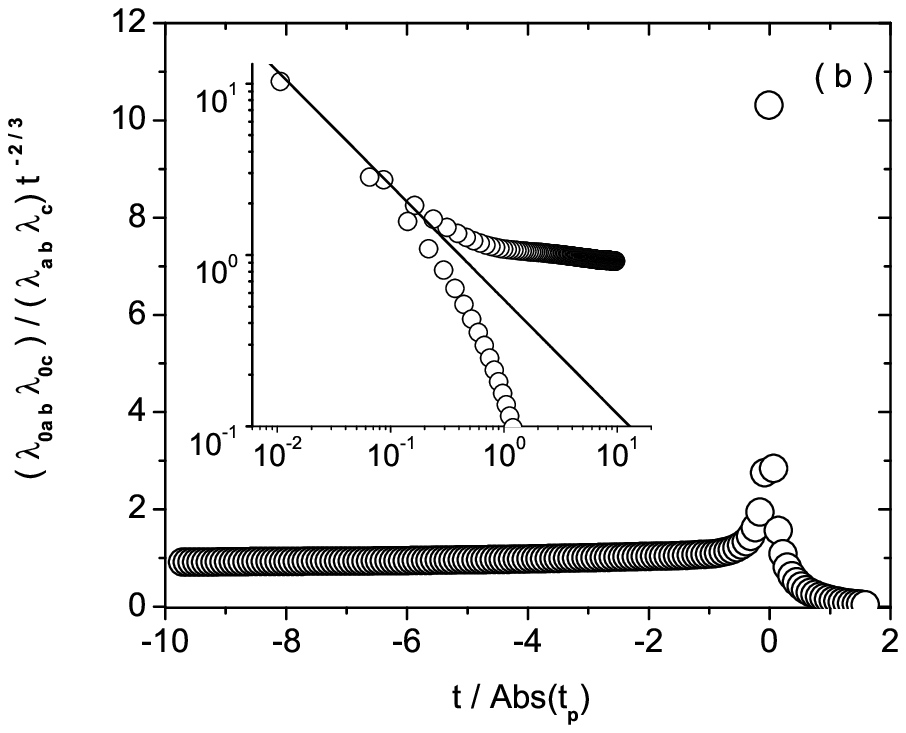}
\caption{Finite size scaling function, (a) $\left| t\right| ^{-\nu
}\left( \lambda _{0ab}/\lambda _{ab}\left( T\right) \right) ^{2}$
versus $t/\left| t_{p_{c}}\right| $ and (b) $\left| t\right|
^{-\nu }\left( \left( \lambda _{0ab}\lambda _{0c}\right) /\left(
\lambda _{ab}\left( T\right) \lambda _{c}\left( T\right) \right)
\right) $ versus $t/\left| t_{p_{ab}}\right| $ derived from the
data shown in Fig.\ref{fig7} for YBa$_{2}$Cu$_{3}$O$_{6.7}$ with
the parameters listed in Table III. The insert shows the log
plot.} \label{fig8}
\end{figure}

To complete the finite size scaling analysis and the evidence for
superconducting domains with diameters $L_{c}$ and $L_{ab}$, we
displayed in Fig.\ref{fig8} the scaling functions $\left| t\right|
^{-\nu }\left( \lambda _{0ab}/\lambda _{ab}\left( T\right) \right)
^{2}$ versus $t/\left| t_{p_{c}}\right| $ and $\left| t\right|
^{-\nu }\left( \left( \lambda _{0ab}\lambda _{0c}\right) /\left(
\lambda _{ab}\left( T\right) \lambda _{c}\left( T\right) \right)
\right) $ versus $t/\left| t_{p_{ab}}\right| $. Together with
Eqs.(\ref{eq10c}) to (\ref{eq10d}) we observe the required
agreement with the behavior of the respective finite size scaling
function. The near coincidence $g_{0c}\simeq g_{0ab}\simeq 0.5$
implies nearly identical shape and boundary condition along the
c-axis and the ab-plane. Moreover, $g_{0c}\simeq 0.5$ is close to
the value in the Bi$_{2}$Sr$_{2}$CaCu$_{2}$O$_{8+\delta }$ films
(Table I). To provide a consistency check of the finite size
scaling analysis we note that, given the estimates for $g_{0k} $,
$\sqrt{\xi _{0i}^{t}\xi _{0j}^{t}}$ and $L_{k}$, the quantity
$g_{0k}\sqrt{\xi _{0i}^{t}\xi _{0j}^{t}}/L_{k}$ is readily
calculated and should be equal to $\left( \lambda _{0i}\lambda
_{0j}\right) /\left( \lambda _{i}\left( T_{c}\right) \lambda
_{j}\left( T_{c}\right) \right) $(Eq.(\ref{eq10e})). Tables II and
III reveal satisfactory agreement.
To summarize, we have shown that the tail in the temperature
dependence of the in-plane and out of plane penetration depth
around $T_{c}$, as observed in the experimental data considered
here, is fully consistent with a finite size effect, arising from
homogeneous nanoscale superconducting domains with diameters
$L_{ab}$ and $L_{c}$. Clearly this finite size effect is not
restricted to the penetration depth but should be visible in other
thermodynamic properties, including the specific heat, as well. In
the specific heat it leads to a rounding of the peak and its
consistency with a finite size effect was established for the data
taken on YBa$_{2}$Cu$_{3}$O$ _{7-\delta }$ high quality single
crystals\cite{book,tsphysB}. In these samples the domain size was
found to range from 300 to 400 \AA. Although the investigations of
Gauzzi \emph{et al.}\cite{gauzzi} on YBa$_{2}$Cu$_{3}$O$_{6.9} $
films with reduced long-range structural order clearly reveal that
the size of the domains depends strongly on the growth conditions,
we established their nanoscale size and their thermodynamic
relevance in a variety of samples.

We thank K. D. Osborn \emph{et al.} for providing their
experimental data.

\end{document}